\documentclass[DIV12]{scrartcl}

\usepackage{hyperref}
\usepackage[utf8]{inputenc}
\usepackage{tabularx}
\usepackage{setspace}
\usepackage{graphicx}
\usepackage{booktabs}
\usepackage{amsmath}
\usepackage{color}
\usepackage{url,cite} 
\usepackage{caption}
\usepackage{verbatim} % for block comment
\usepackage{amssymb}
\usepackage{algorithm}
\usepackage{algorithmic}
\usepackage{tikz}
\usepackage{footnote}

\hypersetup{
	colorlinks=true,
	linkcolor=black,
	citecolor=black,
	urlcolor=black
}

\date{}
\usepackage{authblk}

\usepackage{natbib}
\title{Music Signal Processing Using Vector Product Neural Networks}

\usepackage{fancyhdr}
\begin{document}

\author[1]{Zhe-Cheng Fan\thanks{lambert.fan@mirlab.org}}
\author[2]{Tak-Shing T.~Chan\thanks{takshingchan@citi.sincia.edu.tw}}
\author[2]{Yi-Hsuan Yang\thanks{yang@citi.sincia.edu.tw}}
\author[1]{Jyh-Shing R.~Jang\thanks{jang@mirlab.org}}

\affil[1]{\small Dept.~of Computer Science and Information Engineering, National Taiwan University, Taiwan}
\affil[2]{\small Research Center for Information Technology Innovation, Academia Sinica, Taiwan}

\maketitle
\thispagestyle{fancy} 

\begin{abstract}
We propose a novel neural network model for music signal processing using vector product neurons and dimensionality transformations. Here, the inputs are first mapped from real values into three-dimensional vectors then fed into a three-dimensional vector product neural network where the inputs, outputs, and weights are all three-dimensional values. Next, the final outputs are mapped back to the reals. Two methods for dimensionality transformation are proposed, one via context windows and the other via spectral coloring. Experimental results on the iKala dataset for blind singing voice separation confirm the efficacy of our model.
\bigskip

\end{abstract}

\section{Introduction}
\label{sec:introduction}

\begin{comment}
\begin{figure}[t]
\begin{center}
\includegraphics[width=14cm]{Block_Diagram.jpg}
\caption{Block diagram of the proposed framework} 
\label{fig:block diagram}
\end{center}
\end{figure}
\end{comment}

In recent years, deep learning has become increasingly popular in the music information retrieval (MIR) community. For MIR problems requiring clip-level predictions or frame-by-frame predictions, such as genre classification, music segmentation, onset detection, chord recognition and vocal/non-vocal detection, many existing algorithms are based on convolutional neural networks (CNN) and recurrent neural networks (RNN). For audio regression problems which require an estimate for each time-frequency (t-f) unit over a spectrogram, such as source separation \citep{zhang16taslp}, more algorithms are based on deep neural networks (DNN). This is because for such problems both input and output are matrices of the same size and therefore the neural network cannot involve operations that may reduce the spatial resolution. Existing DNN models for such problems usually consider each t-f unit as a real value and take that as input for the network \citep{huang14ismir,Roma16mirex}. The question we want to address in this paper is whether we can achieve better results by applying some transformation methods to enrich the information for each t-f unit. 

A widely used approach to enrich the information of each t-f unit is to add temporal context \citep{zhang16taslp}. For instance, in addition to the current frame, we add the previous-$k$ and subsequent-$k$ frames to compose a real-valued matrix and take it as the input of the neural network. But in this way, the interaction between different dimensions cannot be well modeled. To address this issue, we find it promising to consider a (2$k$+1)-dimensional neural network \citep{nitta07ijcai}. As a first attempt, we implement this using vector product neural network (VPNN) \citep{nitta93ijcnn}, a three-dimensional neural network that has only been tested on a simple XOR task in the literature. That is to say, $k$ in this work is set to 1 (i.e. considering only the two neighboring frames). Each t-f unit is projected to a three-dimensional vector via this dimensionality transformation method. In VPNN, the input, output, weight and bias of each neuron are all three-dimensional vectors. While the main operation in DNN is matrix multiplication, it is the vector product in VPNN. 

The goal of the paper is three-fold. First, we renovate VPNN with modern optimization techniques \citep{ruder16arxiv} and test it on MIR problems instead of a simple XOR problem. Second, we test and compare the conventional context-enriched DNN structure using real values and the context-enriched VPNN structure for the specific task of blind singing voice separation from monaural recordings, which is a type of source separation problem. Third, we implement the idea of spectral coloring as another way to convert a real-valued matrix to a three-dimensional vector-valued matrix and evaluate this method again for blind singing voice separation. Our experiments confirm the efficacy of VPNN and both dimensionality transformation methods.

\section{Vector Product Neural Network}

In VPNN, the input data, weights, and biases are all three-dimensional vectors. Suppose there is an $L$-intermediate-layer VPNN, the input $\mathbf{z}_i^l$  of activation function in each neuron at the $l$-th layer is:
\begin{equation}
\mathbf{z}_i^l=\sum_{j=1}^{J}\mathbf{w}_{ij}^{l}\times\mathbf{a}_j^{l-1}+\mathbf{b}_{i}^{l},
\end{equation}
where $\times$ denotes vector product, $\mathbf{w}_{ij}^l$ stands for the weight connecting neurons $j$ and $i$ at $l$-th layer, $\mathbf{a}_j^{l-1}$ the input signal coming from neuron $j$ at $(l-1)$-th layer, $\mathbf{b}_{i}^{l}$ the bias of neuron $i$ at $l$-th layer. If $\mathbf{x}$=[$x_1$ $x_2$ $x_3$] and $\mathbf{y}$=[$y_1$ $y_2$ $y_3$], the result of vector product operation is $\mathbf{x}\times\mathbf{y}$ = [$x_2y_3-x_3y_2$, $x_3y_1-x_1y_3$, $x_1y_2-x_2y_1$].
%In other words, each dimension of output is the mathematical combination by calculating values from other dimensions, thereby capturing the possible interaction among the three dimensions. This is not available in typical real-valued neural network models. 
As each element of the output vector receives contributions from all other dimensions, the vector product can capture all possible interactions among the three dimensions. This is not possible with real-valued neural network models

Note that we need to compute a lot of vector products between the layers in VPNN. In order to reduce training time, we propose to reformulate the vector product as matrix multiplication, which is more amenable to GPU acceleration. Suppose there are two vector-valued matrices, $\mathbf{P}$ and $\mathbf{Q}$. Their vector-valued matrix product $\otimes$ can be equivalently written as:
\begin{equation}
\mathbf{P}\otimes\mathbf{Q}=[\mathbf{p}_2\mathbf{q}_3-\mathbf{p}_3\mathbf{q}_2, \mathbf{p}_3\mathbf{q}_1-\mathbf{p}_1\mathbf{q}_3, \mathbf{p}_1\mathbf{q}_2-\mathbf{p}_2\mathbf{q}_1],
\label{eq:mtxMultForVecPdt}
\end{equation}
where $\mathbf{p}_1$, $\mathbf{p}_2$, $\mathbf{p}_3$ are the matrices making up $\mathbf{P}$ and $\mathbf{q}_1$, $\mathbf{q}_2$, $\mathbf{q}_3$ are the matrices making up $\mathbf{Q}$. By applying Eq.~\eqref{eq:mtxMultForVecPdt}, the output at hidden layer $l$ can be defined as:
\begin{equation}
\mathbf{A}^l=\phi(\mathbf{W}^l\otimes\mathbf{A}^{l-1}+\mathbf{B}^l),
\end{equation}
and the output $\mathbf{Y}$ of the VPNN, which is a vector-valued matrix, can be defined as:
\begin{equation}
\mathbf{Y}=\phi(\mathbf{W}^L... \phi(\mathbf{W}^2\otimes\phi(\mathbf{W}^1\otimes\mathbf{A}^0+\mathbf{B}^1)+\mathbf{B}^2)+...\mathbf{B}^L),
\end{equation}
where operator $\otimes$ denotes the vector-valued matrix product mentioned in Eq.~\eqref{eq:mtxMultForVecPdt}. At the $l$-th layer, $\mathbf{A}^l$ is the hidden state, $\mathbf{W}^l$ is weight matrix, and $\mathbf{B}^l$ is bias matrix. All of them are vector-valued matrices. At the first layer, $\mathbf{A}^0$ is the input of the VPNN, consisting of vector-valued data from dimensionality transformation. The function $\phi$ is the sigmoid function. 
%The weight and bias were estimated by back-propagation based on matrix multiplication instead of vector product. 
In order to achieve better performance, modern gradient optimization methods \citep{ruder16arxiv} are implemented in our VPNN. Due to space constraints, we only report our results using Adam, a method for stochastic gradient descent.

\section{Dimensionality Transformation and Objective Function}
\label{sec:proposed work}

In this section, we elaborate two ideas for dimensionality transformation. One is based on adding temporal context described in Section \ref{sec:introduction}. The other one is based on a novel technique called spectral coloring, which associates each t-f unit with a color in the RGB color space. Both ideas yield three-dimensional vectors as the input to VPNN.

\subsection{Context-Windowed Transformation}

To enrich the information for each t-f unit and improve the problem of interaction between different dimensions, we make current frame, previous and subsequent frames as a three-dimensional vector for each t-f unit. For ordinary NN, the input would be three real-valued matrices. For VPNN, the input is a three-dimensional matrix. In other words, the first dimension consists of previous frames, second dimension for current frames and third dimension for subsequent frames. Here we call this model as context-Window Vector Product Neural Network (WVPNN). After feeding the three-dimensional matrix into the VPNN, we get three-dimensional outputs and the second dimension is our predicted result.

\subsection{Spectral Color Transformation}

We can also map each t-f unit from a one-dimensional value into a three-dimensional vector, by using so-called spectral color transform, using for example the hot colormap in Matlab directly as a lookup table, where the forward and inverse maps are both computed by nearest neighbor searches. The hot colormap is associated with a resolution parameter which specifies the length of the colormap. As nearest neighbor interpolation will turn into piecewise linear interpolation when $n$ tends to infinity, we can imitate the hot colormap with infinite resolution by the following piecewise linear functions instead:
\begin{equation}
\mathbf{v}=
\begin{bmatrix}
r \\
g \\
b \\
\end{bmatrix}
=
\begin{bmatrix}
\operatorname{max}(\operatorname{min}(x/n,1),0) \\
\operatorname{max}(\operatorname{min}((x-n)/n,1),0) \\
\operatorname{max}(\operatorname{min}((x-2n)/(1-2n),1),0) \\
\end{bmatrix},
\label{eq:spect2rgb}
\end{equation}
where $\mathbf{v}$ stands for the three-dimensional vector-valued vector, $r$, $g$ and $b$ the R, G, and B values respectively, $x$ the magnitude of each t-f unit, and $n$ a scalar to bias the generation of RGB values. We empirically set $n$ to $0.0938$ in this work. We call this model Colored Vector Product Neural Network (CVPNN). After feeding the RGB values into VPNN, we get RGB values as outputs. Each RGB value is then inversed-mapped to a magnitude at each t-f unit.

% \begin{figure}[t]
% \begin{center}
% \includegraphics[width=16.0cm]{CVPNN_framework.jpg}
% \caption{The framework of CVPNN training.} 
% \label{fig:CVPNN_framework}
% \end{center}
% \end{figure}

\subsection{Target Function and Masking}
\label{sec:target function and masking}

During the training process of singing voice separation, given the predicted vocal spectra $\mathbf{\tilde{Z}_1}$ and predicted music spectra $\mathbf{\tilde{Z}_2}$, together with the original sources $\mathbf{Z_1}$ and $\mathbf{Z_2}$, the objective function $J$ of the WVPNN and CVPNN can be defined as:

\begin{equation}
J=\Arrowvert \mathbf{\tilde{Z}_1}-\mathbf{Z}_1 \Arrowvert^2+\Arrowvert \mathbf{\tilde{Z}_2}-\mathbf{Z}_2 \Arrowvert^2.
\label{eq:J}
\end{equation}

% To make it possible to give different weights to the separation result of the vocal and the instrumental part, in our second experiment we also scale the first term by $c$ and the second term by $1-c$. The objective function then becomes: 
% \begin{equation}
% J_{weighted}=c \Arrowvert \mathbf{\tilde{Z}_1}-\mathbf{Z}_1 \Arrowvert^2+ (1-c)\Arrowvert \mathbf{\tilde{Z}_2}-\mathbf{Z}_2 \Arrowvert^2.
% \label{eq:Jweight}
% \end{equation}
After getting the output from WVPNN and CVPNN, we can obtain the predicted spectra $\mathbf{\tilde{y}_1}$ and $\mathbf{\tilde{y}_2}$. We smooth the results with a time-frequency masking technique called that soft time-frequency mask \citep{huang2015taslp}, and the magnitude spectra of the input frame can be transformed back to the time-domain by inverse STFT with the original phases.

\section{Experiments}
\label{sec:experiment}

The proposed models are evaluated by singing voice separation experiments on the iKala dataset \citep{chan15icassp}.
% which consists of 352 30-second song clips with a sample rate of 44100 Hz. These clips are excerpted from Chinese popular songs performed by professional singers. Each clip is a stereo recording, with one channel for the singing voice and the other for background music.
Only 252 song clips are released as a public set for evaluation. Due to the limitation of GPU memory, we partition the public set into 63 training clips and 189 testing clips. To reduce computation, all clips are downsampled to 16000 Hz. For each song clip, we use STFT to yield magnitude spectra with a 1024-point window and a 256-point hop size. 
%The performance is measured by the Blind Source Separation (BSS) Eval toolbox v3.0 \citep{vincent06taslp}. The overall performance over all the test clips is reported via global NSDR (GNSDR), global SIR (GSIR), and global SAR (GSAR), which are the weighted means of the measures over all clips with a weighting proportional to the length of each clip.
The performance is measured in terms of source to distortion ratio (SDR), source to interferences ratio (SIR), and source to artifact ratio (SAR), as calculated by the blind source separation (BSS) Eval toolbox v3.0 \citep{vincent06taslp}. The overall performance is reported via global NSDR (GNSDR), global SIR (GSIR), and global SAR (GSAR), which are the weighted means of the measures over all clips with a weighting proportional to the length of the clips. Higher numbers mean better performances.

In order to compare the performance of ordinary DNN and CVPNN, we construct two networks that both consist of 3 hidden layers and 512 neurons in each hidden layer, denoted as CVPNN and DNN$_1$, respectively, using Eq.~\eqref{eq:J} as the target function.
%The differences between these two are the number of NN parameters and the dimensionality of each t-f unit. 
The dimensionality of each t-f unit is 1 for ordinary DNN and 3 for CVPNN. A network structure similiar to this DNN$_1$ was used in \citep{Roma16mirex}. As shown in Table~\ref{tab:tab1}, CVPNN performs better than DNN$_1$ in both GNSDR and GSIR. As CVPNN has three times of NN parameters (i.e. number of weights and bias) as compared with DNN$_1$, for fair comparison we further construct an ordinary DNN comprising 3 hidden layers and 1536 neurons in each hidden layer, denoted as DNN$_2$, so that both CVPNN and ordinary DNN have the same number of parameters. Table~\ref{tab:tab1} shows that CVPNN still performs better. 
Besides, we also construct two architectures which have the same number of parameters and t-f units composed of context window size of 3 frames, denoted as WVPNN and DNN$_3$ respectively. Both of them are composed of 3 hidden layers. The number of neurons in each hidden layer is 512 for WVPNN, and 1536 for DNN. The difference of these two is the combination of input frames. The input frames is constructed as a three-dimensional vector-valued matrix for WVPNN and a two-dimensional valued matrix for DNN$_3$. Results in Table~\ref{tab:tab1} show that WVPNN performs better than ordinary DNN$_3$.

\begin{table}[t]
\centering  
\caption{Comparison of ordinary DNN, WVPNN and CVPNN.}
\label{tab:tab1}  

\begin{tabular}{clcccc}  

\hline
Neural Networks.  &Arch.     &Context Window Size  &GNSDR &GSIR &GSAR \\ \hline 
DNN$_1$       &512x3         &1            &8.16 &11.88 &12.11 \\  
DNN$_2$       &1536x3        &1            &8.37 &12.64 &11.82 \\         
CVPNN         &512x3         &1            &8.87 &13.38 &11.37 \\ \hline
DNN$_3$       &1536x3        &3            &8.85 &12.59 &12.52 \\
WVPNN         &512x3         &3            &9.01 &13.82 &11.97 \\       
\hline

\end{tabular}
\end{table}

\section{Conclusion and Future Work}
\label{sec:concl}

In this paper, we propose WVPNN and CVPNN for monaural singing voice separation, using two dimensionality transformation methods. We also propose modern gradient optimization methods on VPNN to attain better performance. Our evaluation shows that both proposed models are better than traditional DNN, with 0.16--0.85 dB GNSDR gain and 1.23--1.94 dB GSIR gain. Future work is to extend these models to CNN and RNN and apply them to other MIR problems, such as genre classification and music segmentation.

\bibliography{paper}

\begin{thebibliography}{9}
\providecommand{\natexlab}[1]{#1}
\providecommand{\url}[1]{\texttt{#1}}
\expandafter\ifx\csname urlstyle\endcsname\relax
  \providecommand{\doi}[1]{doi: #1}\else
  \providecommand{\doi}{doi: \begingroup \urlstyle{rm}\Url}\fi

\bibitem[Chan et~al.(2015)]{chan15icassp}
Tak-Shing Chan et~al.
\newblock Vocal activity informed singing voice separation with the {iKala}
  dataset.
\newblock In \emph{Proc. ICASSP}, pages 718--722, 2015.

\bibitem[Huang et~al.(2014)]{huang14ismir}
Po-Sen Huang et~al.
\newblock Singing-voice separation from monaural recordings using deep
  recurrent neural networks.
\newblock In \emph{Proc. ISMIR}, pages 477--482, 2014.

\bibitem[Huang et~al.(2015)]{huang2015taslp}
Po-Sen Huang et~al.
\newblock Joint optimization of masks and deep recurrent neural networks for
  monaural source separation.
\newblock \emph{IEEE Trans. Audio, Speech and Language Processing}, 23\penalty0
  (12):\penalty0 2136--2147, 2015.

\bibitem[Nitta(1993)]{nitta93ijcnn}
Tohru Nitta.
\newblock A backpropagation algorithm for neural networks based an {3D} vector
  product.
\newblock In \emph{Proc. IJCNN}, pages 589--592, 1993.

\bibitem[Nitta(2007)]{nitta07ijcai}
Tohru Nitta.
\newblock N-dimensional vector neuron.
\newblock In \emph{Proc. IJCAI}, pages 2--7, 2007.

\bibitem[Roma et~al.(2016)]{Roma16mirex}
Gerard Roma et~al.
\newblock Singing voice separation using deep neural networks and f0
  estimation.
\newblock In \emph{http://www.music-ir.org/mirex/abstracts/2016/RSGP1.pdf},
  2016.

\bibitem[Ruder(2016)]{ruder16arxiv}
Sebastian Ruder.
\newblock An overview of gradient descent optimization algorithms.
\newblock \emph{arXiv preprint arXiv:1609.04747}, 2016.

\bibitem[Vincent et~al.(2006)]{vincent06taslp}
Emmanuel Vincent et~al.
\newblock Performance measurement in blind audio source separation.
\newblock \emph{IEEE Trans. Audio, Speech and Language Processing}, 14\penalty0
  (4):\penalty0 1462--1469, July 2006.

\bibitem[Zhang and Wang(2016)]{zhang16taslp}
Xiao-Lei Zhang and DeLiang Wang.
\newblock A deep ensemble learning method for monaural speech separation.
\newblock \emph{IEEE Trans. Audio, Speech and Language Processing}, 24\penalty0
  (5):\penalty0 967--977, 2016.

\end{thebibliography}

\end{document}